# Monitoring electrochemical dynamics through single-molecule imaging of hBN surface emitters in organic solvents

Eveline Mayner[1], Nathan Ronceray[1], Martina Lihter[1,2], Tzu-Heng Chen[1], Kenji Watanabe[3], Takashi Taniguchi[3], Aleksandra Radenovic[1]

[1] *Laboratory of Nanoscale Biology, Institute of Bioengineering Ecole Polytechnique Federale de Lausanne (EPFL), CH-1015 Lausanne, Switzerland*

[2]*Institute of Physics, Zagreb, Croatia*

[3]*Research Center for Materials Nanoarchitectonics, National Institute for Materials Science, Tsukuba, Japan*

## Abstract

**Electrochemical techniques conventionally lack spatial resolution and average local information over an entire electrode. While advancements in spatial resolution have been made through scanning probe methods, monitoring dynamics over large areas is still challenging, and it would be beneficial to be able to decouple the probe from the electrode itself. In this work, we leverage single molecule microscopy to spatiotemporally monitor analyte surface concentrations over a wide area using unmodified hexagonal boron nitride (hBN) in organic solvents. Through a sensing scheme based on redox-active species interactions with fluorescent emitters at the surface of hBN, we observe a linear decrease in the number of emitters under positive voltages applied to a nearby electrode. We find consistent trends in electrode reaction kinetics vs overpotentials between potentiostat-reported currents and optically-read emitter dynamics, showing Tafel slopes greater than 290 mV decade$^{-1}$. Finally, we draw on the capabilities of spectral single molecule localization microscopy (SMLM) to monitor the fluorescent species identity, enabling multiplexed readout. Overall, we show dynamic measurements of analyte concentration gradients at a micrometer-length scale with nanometer-scale depth and precision. Considering the many scalable options for engineering fluorescent emitters with 2D materials, our method holds promise for optically detecting a range of interacting species with unprecedented localization precision.**

# Introduction

Nanomaterials have emerged as promising high-performance sensors, offering exceptional sensitivity, temporal response, selectivity, and robustness[1]. One of the best-studied nanomaterial-based sensors is the nitrogen-vacancy (NV) center in nanodiamonds, which has fluorescence sensitive to temperature[2], pH[3], strain[4], and electric/magnetic field[5–8]. However, precise engineering of NV centers in nanostructures remains challenging and emission is sensitive to the crystallographic orientation and the distance to surface. Alternatively, the atomic flatness and "surface-only" nature of 2D materials alleviates these difficulties while retaining robustness. In this study, we propose the use of hexagonal boron nitride (hBN), a wide band-gap 2D material that has high thermal, mechanical, chemical, and optical stability to measure interfacial reactions in organic solvents[9,10].

Single molecule or spatially resolved electrochemistry typically relies on scanning with a probe (electrochemical atomic force microscopy, scanning electrochemical microscopy, scanning tunneling microscopy, and scanning electrochemical cell microscopy)[11,12]. While scanning probe methods can provide detailed information about single molecules, including the formation of bonds[13–15], they are limited in field of view and prohibit sensing in confinement. High spatiotemporal resolution can be enabled by micro- or nano- arrays but fabrication methods are often intensive, multistep, and expensive[16,17]. Optical methods are an attractive solution to this problem but non-fluorogenic redox pairs are difficult to study. Recently, electrochemical control over Alexa 647 fluorophore blinking was demonstrated[18] and was shown to be beneficial for super-resolution/STORM microscopy[19]. Conversely, in this work, we show that fluorescent emitters enable a local readout of electrochemical reactions occurring in their vicinity. Using label-free opto-electrochemical measurements of an unmodified 2D material, our method retains spatial sensitivity while harnessing a large field of view and fast temporal resolution. These factors are particularly relevant to heterogeneous samples, determining analyte concentration gradients, and dynamically tracking analyte concentrations.

hBN's band gap of ~6 eV, makes it transparent to visible light. Still, defects in the atomic lattice create highly reactive centers and localized intra-bandgap states that absorb and emit light in the visible range[20–23]. While defects in 2D materials can be considered detrimental, in hBN they warrant applications of their own due to their environmental sensitivity: Electrical control of hBN quantum emitters has already demonstrated in vdW heterostructures[24–26] and the ODMR and photoluminescence (PL) of hBN defects were observed to change in response to temperature and magnetic field[27]. Although most experimental work is completed in air or in vacuum, recent works have demonstrated oxygen-plasma related defect fluorescent sensitivity to acidity in water[28], boron vacancy defects' spin relaxation time dependence on paramagnetic environment,[29] and DNA interaction with unmodified and modified hBN[30,31]

Recently, we reported that organic solvents activate single-photon emitters at the surface of unmodified boron nitride[32]. Our current work showcases electrochemical modulations of these emitters, facilitating a spatially resolved readout of the activity of neighboring electrodes. Using spectral SMLM, we monitor the electrochemically-induced optical changes at the solid-liquid interface, which directly report on analyte dynamics including diffusion and reaction kinetics. The methodology relies on the read-out of probes whose position is localized with nanometric resolution and over several micrometer ranges. By varying the potential difference of neighboring electrodes, we modulate the concentrations of analytes interacting with hBN, demonstrating the prospect of sensing trace species, possibly more reaction-specific than standard current tracing with a potentiostat. Several electrode configurations were tested using two- and three- electrode systems. Finding that the electric field could not be the source of modulation, we focused on the alteration of concentrations of species in the solvent and proposed that the emitter's density modulation is based on the modulation of the concentration of a quencher in solution.

## Results and Discussion

Fluorescence was monitored *in-situ* using a spectral SMLM microscope[33] configured with a homebuilt electrochemical cell (**Figure 1a**, details in **Supplementary Figure 1**), enabling high-resolution widefield imaging (see *Materials and Methods*). Three different electrode configurations were used to narrow the mechanism involved. These configurations will be referred to based on the orientation of the electric field applied to the hBN: out-of-plane, in-plane, and stray. While hBN is non-conductive and thus cannot act as an electrode itself, flakes were exfoliated from bulk and placed in proximity to working electrodes (indium tin oxide, i.e. ITO, or titanium) and the electrochemical potential of these electrodes was modulated. The general acquisition scheme (**Figure 1**) is as follows: Image stacks are acquired (typically 15-50 ms per frame) while electrochemical potential is controlled. Two channels are read-out onto the EMCCD chip (spatial and spectral regions) and a uniform region of the flake is isolated. Emitters are localized (see *Materials and Methods)* to provide information on counts, spectra, and residence times. The trajectories of emitters were also determined but their distribution remained isotropic in all field configurations. The example frames show hBN emitters in methanol optically responding to a +1.25 V pulse at the surrounding ITO electrode vs Ag/AgCl. The change in number of emitters can easily be observed (**Figure 1b-c**) at high changes in applied electrochemical potential ($\Psi_{app}=\Psi_w-\Psi_{ref}$: the potential at the working electrode versus the potential at the reference electrode).

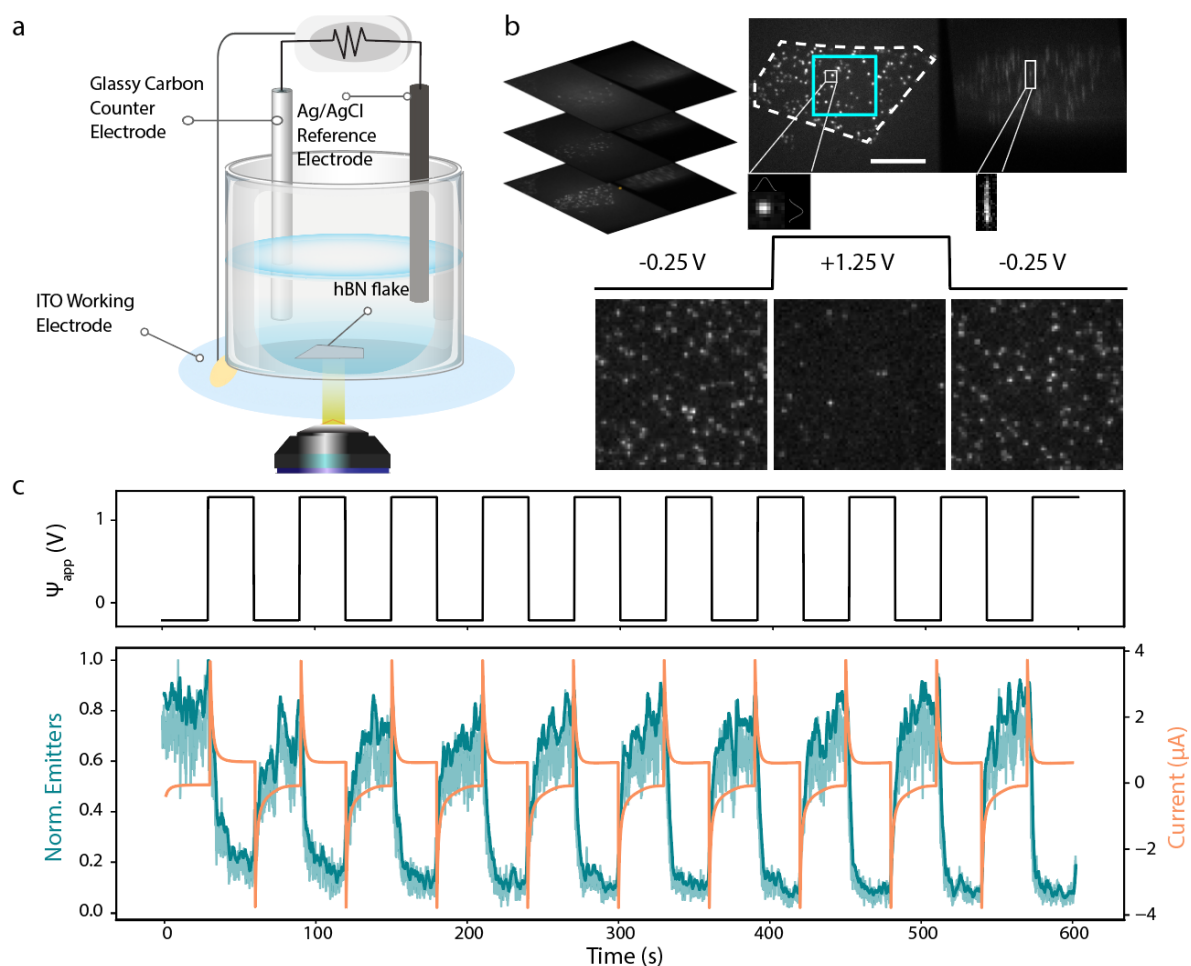

**Figure 1: Acquisition scheme for opto-electrochemical measurements using hBN in organic solvents.**

**a)** A simplified schematic of the opto-electrochemical setup in the out-of-plane three-electrode configuration. An inverted widefield microscope with a high NA objective is used to observe ensembles of hBN solvent emitters with the electrochemical chamber mounted above. **b)** Frames of an hBN flake are acquired while changing the electrochemical potential of the surrounding ITO electrode. Only uniform areas (marked in teal) are used for detection and localization of emitters, meaning edges and creases of hBN flakes are excluded. The scale bar is 5 micrometers. The emitted light is split into two channels: spatial (2D gaussian emitters) and spectral (spread emission after passing through prism). The spatial channel shows the modulation of active emitters by the electrochemical bias in methanol. Several prototypical frames acquired with 30 ms frame rate are shown while switching between -0.25 V and +1.25 V vs Ag/AgCl, demonstrating the clear decrease in the number of active emitters when a positive voltage is applied. **c)** In the spatial channel, emitter counts are time-correlated with the applied waveform and are used to show reversibility and quantify electrochemical effect. The response to 30 second pulses can be seen. Raw data is shown with 0.5 transparency and data smoothed with rolling average of 20 frames is shown in dark teal. The corresponding current trace is shown in orange.

## Out-of-Plane Measurements

In the out-of-plane electrochemical configuration, thick (several tens to few hundred layers) hBN flakes were exfoliated onto glass coverslips coated with a layer of indium-tin-oxide (ITO), shown in **Figure 2a**. The thickness of the flake was not controlled but can be roughly gauged by optical contrast. Flake thickness showed no impact on measurement, as is intuitive with a surface mediated mechanism. The ITO (~70 nm thick) served as the working electrode in a three-electrode configuration while remaining transparent enough to retain a high photon capture. The glassy carbon counter electrode and leakless Ag/AgCl reference electrode were suspended in the solvent above the focal plane. Emitters were excited using a 561 nm laser (power density ~1.6 kW cm$^{-2}$ unless otherwise specified) and the fluorescent signal was collected through the same objective. A potentiostat was used to apply the potential to ITO working electrode vs. Ag/AgCl reference ($\Psi_{app}$).

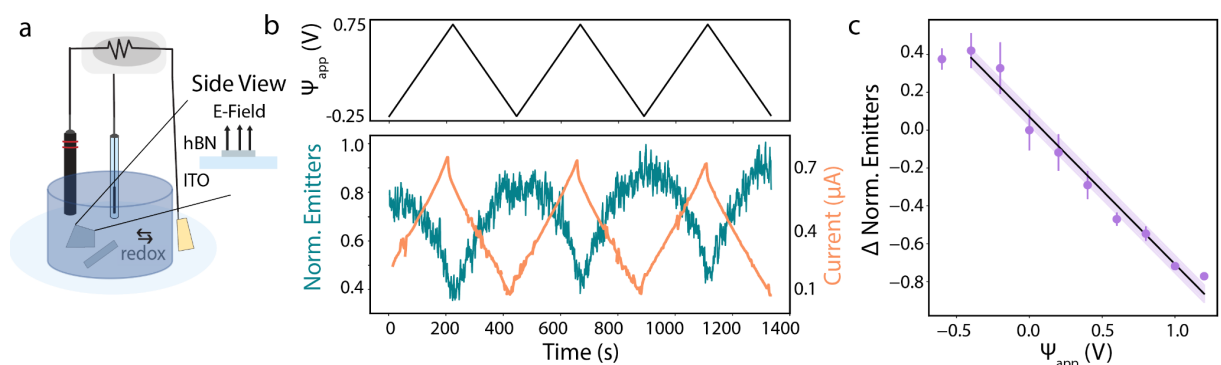

**Figure 2: Out-of-plane opto-electrochemical measurement with varying applied potential. a)** Simplified schematic of the three-electrode out-of-plane measurement configuration with the inset side view showing the field orientation relative to the flake. hBN flakes are transferred to the ITO-coated coverslip which is in contact with the solvent above. A glassy carbon electrode serves as a counter electrode and a thin leakless Ag/AgCl electrode as the reference electrode. **b)** The top panel shows the potential vs Ag/AgCl ($\Psi_{app}$) applied to the ITO working electrode with a triangular waveform for over 20 minutes in acetonitrile. The bottom panel shows the resulting normalized emitter counts (teal) and current (orange). **c)** The change in number of emitters was quantified for various $\Psi_{app}$ using different flakes in methanol (full data in **Supplementary Figure 2**). The change was always characterized relative to that specific experiment's emitter counts at $\Psi_{app}$ =0 V. A linear decrease in the number of emitters per frame is seen at increasingly positive voltages from a $\Psi_{app}$ of -0.4 V, which is within the inert region of the electrode in methanol. Standard errors for each voltage are shown as bars and for the linear fit as shading.

Using the spatial channel, the number of active emitters per frame was monitored while varying the potential applied to the ITO working electrode vs Ag/AgCl reference. Voltage waveforms varied in shape (triangular, square), magnitude, and time duration. An example of a triangular waveform applied in acetonitrile is shown in the top panel of **Figure 2b** and the corresponding current in the bottom panel. The impact of this potential can be seen on the normalized number of emitters per frame, which follows inversely the shape and magnitude of the current. The traces also show that emitter quenching during cycling is reversible, providing evidence that the involved redox reactions are reversible, although the forward and backward reactions may occur at different rates. Square voltage pulses were also used in methanol (data in **Supplementary Figure 2**) to quantify the change in emitters against changing potential in the range of $\Psi_{app}$ = -0.6 V to 1.2 V. The displayed plot is derived from the response of five different flakes to different patterns of square pulses. At increasingly positive potentials and higher anodic currents, the number of active emitters decreases while at increasing negative potentials, the number of active emitters increases. This inverse relationship can be attributed to a change in concentration of species involved in activating or quenching the hBN emitters. The slight decrease at $\Psi_{app}$ = -0.6 V in methanol (**Figure 2c**) can be understood due to the instability of ITO at this potential in organic solvents and higher magnitude negative potentials were not probed in this configuration for this reason. This limited stability has been previously documented[34,35] and the stable range for various organic solvents used in this paper were determined by cyclic voltammetry (CV, available in **Supplementary Figure 3**).

In the out-of-plane orientation, the origin of the optical modulation was elucidated by passivating the surface with a thin insulating layer of aluminum oxide (100 nm). While the electric field remained consistent in orientation and magnitude, this passivation effectively suppressed the analyte redox reaction leading to the suppression of the response of emitters. Therefore, we could conclude that the emitter response originates from chemical species diffusing from the electrodes, resulting in concentration changes during cycling. The thin film characterization via ellipsometry and voltammetry results are available in **Supplementary Figure 4**.

The dependence of the signal on the distance to the electrode was first examined using a "stray-field" configuration wherein the ITO electrode was replaced by a thin titanium electrode patterned onto a glass slide (**Supplementary Discussion on Stray-Fields, Supplementary Figure 5**). The signal modulation was consistent with the out-of-plane measurements, despite the change in orientation of the field and change in electrode, while the magnitude of effect was influenced by distance to the electrodes. This micrometers distance dependence is a useful characteristic for a high spatial resolution sensor. Although we probed inside the working potential window of ITO, the change of electrode material also demonstrates that the decrease in emitter activity near positive electrodes was not due ITO electrode deterioration— ITO is an n-type semiconductor so at high positive potentials, its conductivity and stability decreases[36].

## Examining Reaction Kinetics and Spectra

We then turned to the applicability of our optical method to report on dynamic changes, specifically focusing on anodic reaction kinetics. Since the Faradic current in an electrochemical system is by definition proportional to the reaction rate at the electrode surface by **Equation 1**

$$I = -nFAr \qquad \text{Eq 1}$$

where *I* is the current, *n* is the number of electrons transferred, *A* is the area, *F* is the Faraday constant, and *r* is the reaction rate. This means that for an anodic reaction the current density is described by **Equation 2**

$$i_a = -nFk_{ox}[R_0] \,^{37} \qquad \text{Eq 2}$$

$i_a$ is the anodic current density in A/m$^2$, $k_{ox}$ is the oxidation rate constant, and $R_0$ is the surface concentration of the reductant. Since electrochemical currents are potential-dependent, it follows that the rate constant is also potential-dependent. Although the reaction would occur at the ITO electrode, optical modulation of the hBN emitters provides another means of determining reaction rate constants, given that our analyte reaction occurs proportionally to the bulk current. In our case we can analyze ensemble emitter dynamics while applying pulses of positive potentials to the ITO electrode.

While Tafel analysis is most commonly applied in the contexts of electrocatalysts, it is applicable to any electrochemical reaction to inform on rate limiting steps[37]. However, one should note that Tafel analysis does have significant limitations. For example, it cannot differentiate two mechanisms that share the same expected Tafel slope[38,39]. In our case, the slopes can be influenced by any step of the reaction on ITO: The diffusion of the reactant to the electrode or the reaction at electrode surface, which itself can be complex and rate limiting. At the anode we are considering the oxidation of water (discussed in the **Proposed Mechanism and the Role of Water and H+** section), which is commonly accepted as a reaction with a rate limited by diffusion.

The Tafel slope is derived at high anodic overpotentials from the simplified Butler-Volmer equation, which leads to a simple relationship between the overpotential, η, and the current by **Equation 3**

$$\eta = a + b * log_{10}(i) \qquad \text{Eq 3}$$

where *b* is the Tafel slope. The overpotential is defined as the difference between the electrode potential, *E*, and the standard potential, $E_0'$ (the half-cell potential). Applied potential was used instead of overpotential, but since $E_0'$ is a constant it does not impact the derived slope.

Long pulses were applied to the working electrode in the three-electrode out-of-plane configuration at increasingly positive potentials for seven different potentials using an acquisition time of 15 ms over a 9x9 µm$^2$ area (example trace in **Figure 3a**). The reaction kinetics derived from the potentiostat-reported current were compared to those derived using the emitter temporal dynamics (**Figure 3b**). Smaller Tafel slopes mean that the reaction follows faster kinetic processes because the same currents can be achieved at lower overpotential. Although Tafel slopes generally relate to reaction kinetics rather than mass transport, differentiating between a slow reaction and a diffusion limited reaction can be difficult. For example, the theoretical Tafel slope of the well-known HER reaction is 30 mV dec$^{-1}$ but experimentally reported slopes can reach over 290 mV dec$^{-1}$ [37,40] because the reaction is complicated, faces significant barriers, and/or has rates limited by diffusion.

The optically-derived slope was determined by fitting the exponential decay behavior in emitter counts to first order reactions and determining the kinetic rate constants. In **Figure 3c** this decay is shown averaged from each pulse in **Figure 3a**. The reaction order was determined to be first order by comparing the R$^2$ goodness of fit for first and second order modeling (see **Supplementary Discussion on Kinetics**, **Supplementary Figure 6**). The consistency in magnitude between the optically- and electrochemically-derived slopes validates the optical method for monitoring an electrochemical process part of the current and further that the analyte reaction faces significant barriers.

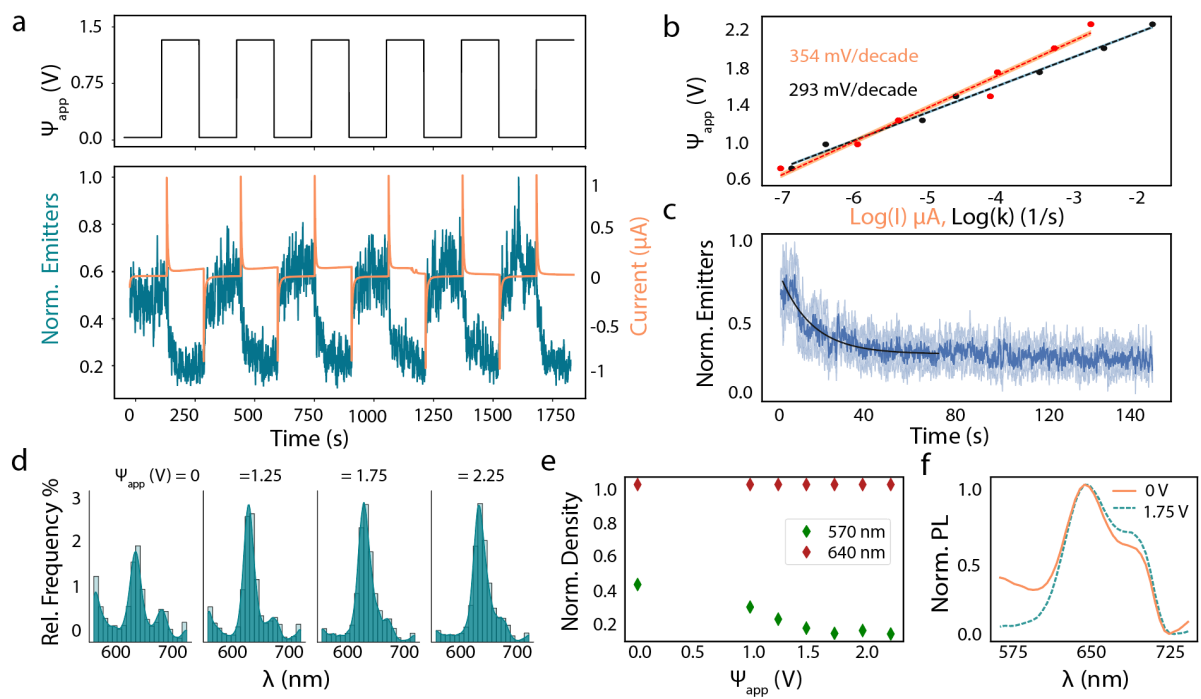

**Figure 3: Out-of-plane localization response characteristics. a)** Long square pulses (150s) were applied cyclically in acetonitrile while the localizations were recorded. The current response to $\Psi_{app}$ = 1.5 V pulses is shown in orange while the response of the normalized number of emitters, is shown below in teal. **b)** The Tafel plot obtained from steady state currents recorded during the pulsing experiments shown in part **a**. The Tafel slope derived from a linear fit of these seven points is indicated by the black dashed line and standard deviation of fit in light blue. The orange dashed line indicates linear fit with shadowed standard deviation for the emitter-derived slope. **c)** An example decay of emitters in response to a positive pulse averaged from the 5 pulses shown in **a**. The dark blue line shows the mean number with a light blue cloud of the standard deviation. **d)** The kernel density estimation (KDE) overlaid on the histogram of spectral peaks of single emitters at various voltages in acetonitrile, obtained using our sSMLM. The characteristic main peak around 640 nm consistently leads. **e)** From the histogram two main peaks (570 nm and 640 nm) relative prominences are plotted. The main group at 640 is constant stray spectral groups are reduced at increasing voltages. **f)** The average sSMLM-obtained spectra of all emitters at Ψapp=1.75 V in acetonitrile. From fitting the averaged PL spectra at $\Psi_{app}$ =1.75 V we find the zero phonon line to be 643 nm and the phonon sideband to be 687 nm.

The spectra of single emitters were used to monitor the identity of the emitters at the same seven increasingly positive potentials in acetonitrile. The acquisition time was increased to 50 ms to increase SNR and to record the spectral channel. Spectra was obtained using the spectral channel of our sSMLM setup (**Figure 1**). The histogram and kernel density estimation (KDE) of spectra from single emitters are seen in **Figure 3d** at various voltages (20 bins, 0.75 binwidth smoothing). Plotting the density of different spectral groups from the histograms at increasing potentials (**Figure 3e**) shows that the group with zero-phonon line (ZPL) around 640 nm is consistently dominant but that there is a pronounced effect of voltage on the spectral group with a peak around 570 nm, a group that has previously been observed on plasma treated and CVD-grown hBN in water[33]. The same purging effect on stray emitters is seen in in-plane at high potentials (**Supplementary Figure 7**). With the observation of several types of emitters with distinct responses to voltage in the same field of view, we demonstrate the opportunity for multiplexing in optical electrochemistry. A video of switching between +/-1.5 V in acetonitrile used for spectral analysis (acquisition: 50 ms) is available as **Supplementary Video 1**.

Examining the spectral channel also revealed that the Stark effect does not play a significant role on the main emitter population: **Figure 3f,** shows the result of averaging individual spectra at $\Psi_{app}$= 0 V and 1.75 V. At 1.75 V, averaging essentially erases the presence of other edge groups and plotting the ZPL and phonon sideband (PSB) shows no significant change (**Supplementary Figure 8**). Finally, by using sub-diffraction localization and single molecule tracking techniques, the residence times of individual emitters was determined. Groups of long-lasting and short-lasting emitters emerged from the bulk at positive potentials showing groups with distinct responses to potential (**Supplementary Figure 9**), demonstrating in residence time another opportunity for monitoring of distinct emitters in the same field of view.

## In-Plane Measurements Demonstrate a Spatially Responsive Electrochemical Sensor

Having characterized the emitter response using an isotropically surrounding electrode, we turned to an in-plane electrode configuration wherein the distance from the active electrode surface could be controlled. We patterned two titanium electrodes onto glass coverslips (depicted in **Figure 4a**) using a custom silicon/silicon nitride shadow mask and electron-beam evaporation (see *Materials and Methods* for details). hBN flakes were then deterministically placed in between these electrodes. The electrodes were oriented such that there was a blank strip 20 to 40 micrometers wide, bisecting the titanium layer and creating two large pads. The top titanium electrode was connected to the working electrode and the bottom electrode to the counter/reference electrode. With the flake in symmetric contact with both electrodes, electrode polarization was switched in time. The total emitter counts responded the same way in response to pulses, regardless of electrode polarization. In acetonitrile, emitter counts spiked initially at each pulse and fell to an equilibrium during the 90 second pulses (example in **Figure 4b**), this can be understood as a capacitive spike occurring in the movement of ions in solution. Other contact orientations were also tested and are reported in **Supplementary Figure 10** and shows contact plays no role except for due to the inherent distance from electrode changes.

Emitter density was found to be higher near the negatively charged electrode (**Figure 4c-d**) and lower near the positively charged electrode, indicating that we can simultaneously monitor both redox half reactions in the same field of view. Concentration gradients can even be seen changing in time, with the change in analyte profile in acetonitrile changing rapidly over ~1 second (**Figure 4e**). To establish the generality and examine the influence of solvent properties, we then repeated measurements using ethanol, isopropyl alcohol, and octanol (concentration profiles over time for other solvents available in **Supplementary Figure 11**). The center of mass (COM), defined as the average x and y position of all localizations in that frame, was tracked while the electrode polarization was switched for the three cycles (**Figure 4f**). The switching region of each solvent was fit linearly (**Figure 4g**). Finally, the reciprocal of the slope was plotted as a function of the viscosity of the solvent (**Figure 4h**). As is understood by Fick's Law, the rate of diffusion is directly proportional to diffusivity (the inverse of viscosity) so increasing solvent viscosity results in a slower diffusion. Because the COM switching rate was found to be inversely related to the viscosity, it follows that either the rate of movement from the electrode to the flake of the redox analyte was limited by diffusion or that the rate-limiting step in the reaction at the electrode was limited by diffusion. Since we observed high Tafel slopes between current- and emitter- derived kinetics, a hypothesis of diffusion limited reaction remains reasonable, albeit likely involving multiple factors.

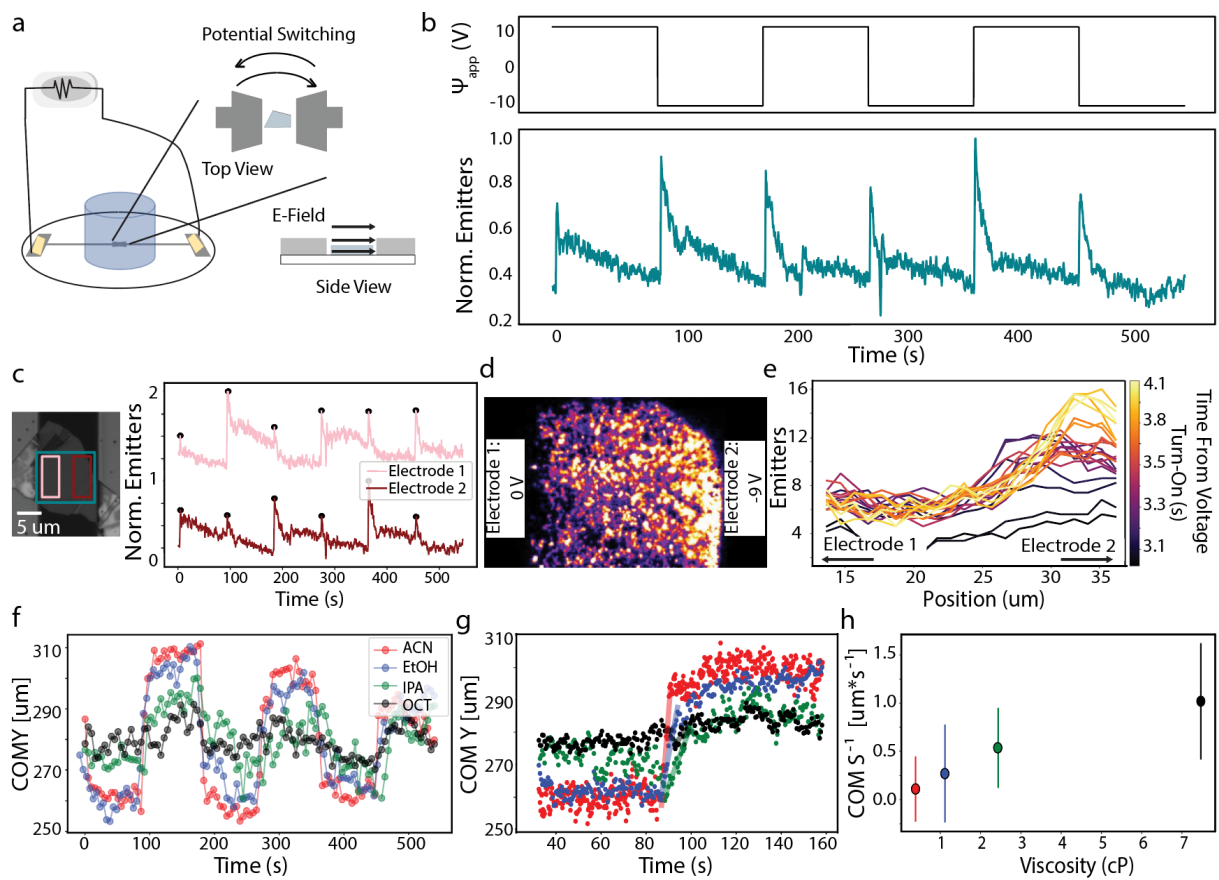

**Figure 4: In-plane experiments show spatial modulation. a)** Simplified schematic of in-plane two-electrode configuration with the inset side view showing the field orientation relative to the flake. The patterned titanium electrodes are connected to the potentiostat via copper tape contacts that are isolated from the solvent. The top electrode served as the working electrode and bottom as counter/reference. **b)** The potential was varied at the working electrode and the normalized number of localizations for the entire flake region is reported in teal over time (50 ms frame rate) for this example trace in acetonitrile. **c)** The optical microscope image shows an example sample wherein a flake was transferred between two titanium electrodes. The localizations are analyzed by region wherein the pink area ("Electrode 1") is near the working electrode and the red area ("Electrode 2") is near the counter/reference electrode. Normalized localizations are then reported with an offset of 1. **d)** 50 frames were stacked together from one electrode polarization to show how the density of emitters visibly changes over the distance of the flake in response to the concentration of the analyte. The video of the switching occurring is available as **Supplementary Video 2**. **e)** In acetonitrile the concentration gradient after the voltage is applied in the first cycle can be seen building over time. The time is reported from the time of the application of the voltage and is shown with inferno color mapping (colorbar on the right). There is a delay from the onset and then the concentration changes over the course of ~1 second. The position is reported as distance from the top of the frame in micrometers and the time is from the onset of the experiment (the first cycle). **f)** The center of mass (COM) of the emitter localizations can be passed between the two electrodes by switching the polarization (as in part b) for acetonitrile (ACN), ethanol (EtOH), isopropyl alcohol (IPA) and octanol (OCT). Here a position point is shown every 5 seconds. **g)** The region corresponding to a switch in polarizations, is averaged for the three pulses and the slope is determined for this linear switching region, corresponding to speed of switching. **h)** The reciprocal of the COM slope (COM S)$^{-1}$ was plotted against the viscosity in centipoise (cP) for these four solvents.

## Proposed Mechanism and the Role of Water and H+

Through the testing of different electrode configurations, we were able to reject an emitter modulation mechanism based on electric field or charge transfer, and to propose a redox-active analyte-based quenching mechanism. Because the modulation is consistent across different solvents and the onset of the effect on emitters is not correlated with the onset potential of a reaction of the bulk solvents (**Supplementary Figure 2**), we consider that there is a common trace species present in the solvents which is modulating the emitter density by quenching. Based on our observation we propose that the quencher may be H+, originating from the oxidation of reactive trace water existing in the solvent. We have also demonstrated the existence of two spectral groups. One group, centered around ~570 nm, has been previously observed in oxygen plasma treated hBN imaged in the presence of water[33], and exhibits notably rapid quenching under increasing potential (**Figure 3d**).

The proposed proton-based mechanism was demonstrated by the deliberate introduction of water and H+ (in the form of HCl) to otherwise pure organic solvents (**Supplementary Figure 13**). Our work shows that as the concentration of water increased, the number of active emitters dropped. Removing water traces with a molecular sieve also removed the effect of voltage cycling at potentials that had a strong impact on "wet" acetonitrile (**Supplementary Figure 14**), indicating that water plays an important role in the modulating reaction mechanism. We also showed that when introducing H+ through HCl, the density of emitters fell more than 25x faster than through introducing water (**Supplementary Figure 13**). This not only indicates the proposed mechanism but shows that fluorescent emitters in our system are highly sensitive to H+ content in organic solvents. Measuring H+ content would be highly relevant in methanol fuel cells[41,42] and using our spatially sensitive system, the concentration could be monitored as a function of distance from electrodes in time.

As is schematically illustrated in **Supplementary Figure 15**, we propose that the emitter density is modulated by the electrochemical reduction/oxidation of a quencher at the two electrodes in the field of view. Although water oxidation typically occurs at higher overpotentials and the concentration of water would be low, (coming from initial presence in purchased solvents[43] and exposure to environment), mixing water with acetonitrile has been shown to create water clusters with increased susceptibility to oxidation due to the modification of its molecular aggregation behavior[44]. Water is a known quencher of red emitting dyes[45] and it has been suggested that this is due to excited state proton transfer from water to emitter[46,47]. Given this adventitious source of water and H+, the mechanism of quenching could be understood by proton quenching of the hBN-solvent emitter. However, in terms of modulating mechanisms, we acknowledge other possible routes involving reactions of the solvent itself and discuss them in **Supplementary Discussion on Mechanisms**. While our findings leave room for more thorough investigations into the quenching mechanism and emitter identity, we nonetheless clearly demonstrate that the quenching is triggered by electrochemical reactions, providing an optical method for monitoring analyte species which can be applied to other systems with modified 2D materials.

## Conclusion

In this article we describe the use of single molecule fluorescence microscopy with unmodified hBN to optically monitor electrochemical reactions and analyte concentrations in time. We also demonstrate the unique ability for local analyte concentration and diffusivity measurements in confinement and locations a scanning probe cannot access. hBN emitters were shown to report on electrochemical reactions occurring in their vicinity as well as concentrations of water and H+. The electrochemical measurement configurations herein presented allowed us to narrow down the mechanism to an electrochemical modulation, ruling out other possible mechanisms such as modulation by charge injection or electric field/orientation. The spatial resolution inherent in the surface-only nature of the hBN and the localization microscopy scheme make it a favorable method for sensing reactions with high spatial specificity, with nanometers resolution in z and micrometers in x and y. The temporal resolution also enables us to report on changes in concentration over tens of milliseconds. With these attributes it is possible to monitor diffusion and the buildup of a concentration gradient in time. Our experiments clearly demonstrate the benefits of integrated opto-electrochemistry. Specifically, we show spatially resolved electrochemical measurements with a large field of view, an approach that holds promise for probing electrochemistry in nonuniform materials and in confinement. Additionally, the work highlights the advantages of using 2D materials, which offer additional engineering opportunities via defect creation or surface modification, facilitating the monitoring of other specific reactions.

## Materials and Methods

### Out-of-Plane Sample Preparation

ITO-coated coverslips (70-100 Ohms Sq$^{-1}$) 25mm diameter x 0.17mm (#1.5 coverslips) thick were purchased from Diamond Coatings LTD. The coverslip is cleaned by sonication in acetone and IPA and finally with oxygen plasma for 90 seconds with 100 watts. High quality hBN flakes were gently exfoliated from bulk crystals[48] using low-adhesion Nitto tape and transferred onto the ITO coverslips. The thickness of the flake was not controlled but could be roughly gauged by optical contrast. It showed no impact on measurement, as is intuitive with a surface mediated mechanism.

### Electrochemical Methods

An electrochemical cell compatible with single molecule localization microscopy was designed for 25 mm coverslip with three possible configurations, as named for the orientation of the electric field relative to the hBN flake: out-of-plane, in-plane, stray-field. The cell was then machined out of PEEK, which is highly resistant to organic solvents. A 3D rendering of the cell is available in **Supplementary Figure 1**.

Cyclic and pulse voltammetry experiments were carried out using a PalmSens4 instrument configured in a three-electrode (out-of-plane, stray-field) or two-electrode (in-plane) configuration. PTFE encapsulated glassy carbon and ITO were used as the counter and working electrodes, respectively. An Ag/AgCl leakless electrode suitable for organic solvents (Alvatek) was used as the pseudo reference electrode. Measurements were carried out at room temperature and any openings to the environment were covered but not sealed in an airtight manner from the environment. A new cell, which had every opening (for electrodes and viewing windows) sealed from the environment was made for the anhydrous experiments shown in **Supplementary Figure 14**. Scan rates of cyclic voltammetry varied from 7.5 to 50 mV s$^{-1}$. The potentials are all reported against the leakless Ag/AgCl electrode in the three-electrode configuration. In the two-electrode configuration the counter and reference electrode were the same titanium electrode.

### In-Plane + Stray-Field Sample Preparation

Electrodes were patterned using silicon/silicon nitride (Si/500 nm Si$_3$N$_4$) stencil masks and evaporation of titanium. The stencil mask was prepared using photolithography (positive photolithography with reactive ion etching) and subsequent KOH etching. The distance between electrodes was retained by the silicon nitride membrane, which remains after isotropic KOH etching. Various electrode spacings were tested by varying this membrane width.

Glass coverslips were cleaned with acetone, IPA, and oxygen plasma before metal patterning with E-beam evaporation. Various electrode heights (100-250 nm) and spacings (10-100 nm) were tested. All wafer fabrication and evaporation processes were done in the clean room.

To place hBN flakes in between (in-plane) or adjacent to (stray-field) patterned electrodes, a dry transfer method with a home-built transfer platform is used. A stamp consisting of a thin film of polypropylene carbonate (PPC) spin coated on a mound of polydimethylsiloxane (PDMS) mechanically stabilized by glass slide is attached to a micromanipulator under a microscope. A $Si_3N_4$/Si substrate with hBN to be transferred, is fixed onto a transfer stage and the desired hBN flake is identified using optical microscopy and positioned with a micromanipulator. Any PPC residue is then washed using acetone and IPA. The mask lithography pattern and an optical image of an example coverslip with transferred hBN between titanium electrodes is available in **Supplementary Figure 16**.

## ITO Surface Passivation

The ITO coverslips were passivated using atomic layer deposition (ALD) of 100 nm of alumina oxide ($Al_2O_3$). Characterization of the thin film was done using ellipsometry in 8 locations in the center 8 mm of the coverslip surface, the area in contact with the organic solvent during imaging (available in **Supplementary Figure 3**). A 3 mm diameter circle adjacent to the edge of the coverslip was isolated from coating during ALD via Kapton tape which was then used for contacting the ITO to the potentiostat via the same silver-tape-cover wire to potentiostat configuration as previously described.

## Spectral Single Molecule Localization Microscopy

The sample was excited using a 561 nm laser (Monolithic Laser Combiner 400B, Agilent Technologies) collimated and focused on the back focal plane of a high-numerical aperture oil-immersion objective (Olympus TIRFM 100x, NA: 1.45). Excitation power was controlled on a range from 10 to 80 mW over a ~2.5 x $10^3$ μm$^2$ illumination spot resulting in a power density of 0.4-3.2 kW cm$^{-2}$. The sample was mounted in a PEEK electrochemical chamber placed on a piezoelectric scanner (Nano-Drive, MadCityLabs). Fluorescent emission is collected by the same objective and separated from excitation using the dichroic and band pass emission filter (ZT488/561rpc-UF1 and ZET488/561m, Chroma) and finally projected on an EMCCD camera (Andor iXon Life 897) with electron multiplying gain of 150. Exposure times ranged from 15 to 50 ms and stacks from 3 to 114 thousand frames.

Spectral SMLM is performed by splitting the emission via a beam splitter into two paths described in detail previously[33] and summarized briefly here. Path 1, the spatial path travels directly to the EMCCD while path 2 passes through an equilateral calcium fluoride ($CaF_2$) prism which spreads the light (spectral path). Acquired image stacks from sSMLM were processed using ThunderSTORM[49]. Emitter filtering and counting were performed using ThunderSTORM, omitting localizations that had intensity less than 300 photons or 30<$\sigma_{PSF}$<250 nm.

For spectral processing only emitters with intensity greater than 350 photons were used. The spectral channel was related to the spatial channel and wavelength via a facile matrix transform($x_{SPEC}$,$y_{SPEC}$) = A × ($x_{LOC}$, $y_{LOC}$) + B where A is a 2 × 2 matrix and B is a column vector. The position of different wavelengths (B vector) was calibrated using broadband beads with narrow bandpass filters. An example bead calibration is shown in **Supplementary Figure 17**. The single molecule spectra were determined by the peak intensity in the box of 6 x 40 pixels centered at the transformed emitter position. In single molecule spectra, the highest intensity position corresponds to the peak wavelength of the emitter and the relative densities of each of these populations was determined using the KDE function of Python library Seaborn with 20 bins and 0.75 binwidth smoothing (**Figure 3d**). Smoothed histograms with spectral SMLM data has previously been used to display population distributions of single molecule spectra[50]. Average ensemble sSMLM spectra (shown in **Figure 3f**) were normalized and fit using two Lorenztians corresponding to ZPL and PSB using the Python 3.7 package LMFIT42[51]. The consistency in wavelength between the spectral population with the highest density and the averaged ensembled fit ZPL also shows the validity of the single molecule approach.

## Chemicals

All the chemicals were purchased with high purity grade. No effect of supplier on hBN fluorescence activation was observed.

## Acknowledgements

E.M. and A.R. would like to thank Professor Kevin Sivula (LIMNO, EPFL) and Professor Jacques-E. Moser (EPFL) for invaluable insights on mechanism and advice on experimental design in the fields of electrochemistry and photochemistry, respectively. E.M. N.R., M.L. T.-H.C. and A.R. acknowledge funding from the European Research Council (grant 101020445—2D-LIQUID), K.W. and T.T. acknowledge support from JSPS KAKENHI (grant nos. 20H00354, 21H05233 and 23H02052) and World Premier International Research Center Initiative (WPI), MEXT, Japan.

# Supplementary Information for "Monitoring electrochemical dynamics through single-molecule imaging of hBN surface emitters in organic solvents"

## Contents

# Supplementary Figures and Discussions

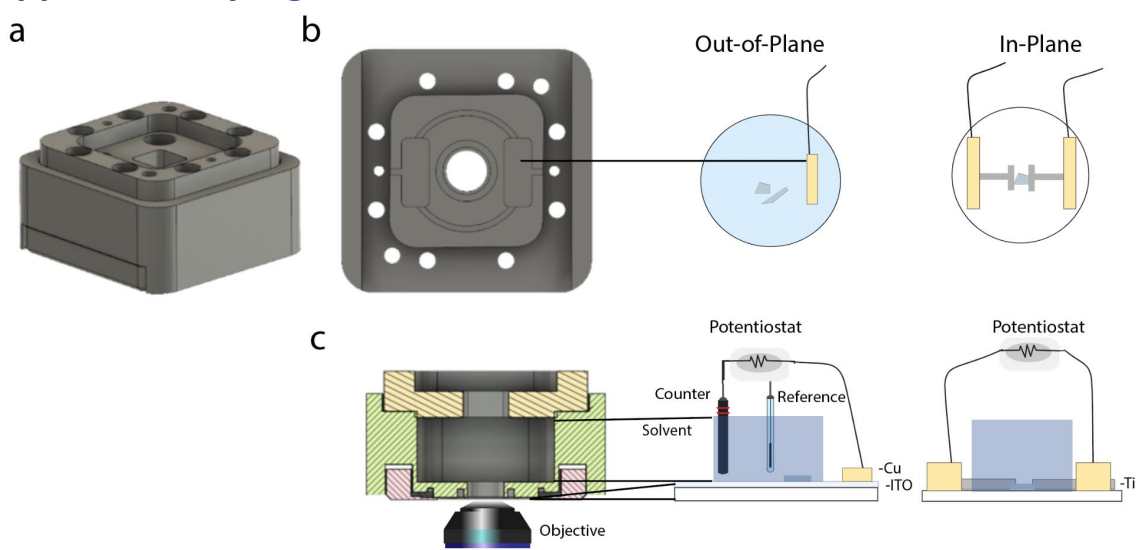

**Supplementary Figure 1: 3D rendering of the electrochemical cell. a)** The electrochemical cell compatible with our high-resolution imaging microscope is comprised of three parts that are screwed together for a tight seal. The view from a 45-degree angle shows the assembly. The top viewing window and the counter electrode port can be seen. There is also a port for the reference electrode out of view. **b)** The middle part, viewed from the bottom, shows how the electrical contact between working ITO electrode (for out-of-plane configuration) or titanium electrodes (for in-plane configuration) are isolated from the fluid in the middle chamber via an O-ring around the center bottom excitation/emission window. The fluid chamber also houses the reference and counter electrodes **c)** A cross-section of the assembled three parts, viewed from the side, illustrates how the chamber facilitates a high-resolution imaging scheme on an inverted microscope, as the objective lens makes contact with the coverslip from below via oil. Excitation and emission are collected through the same objective.

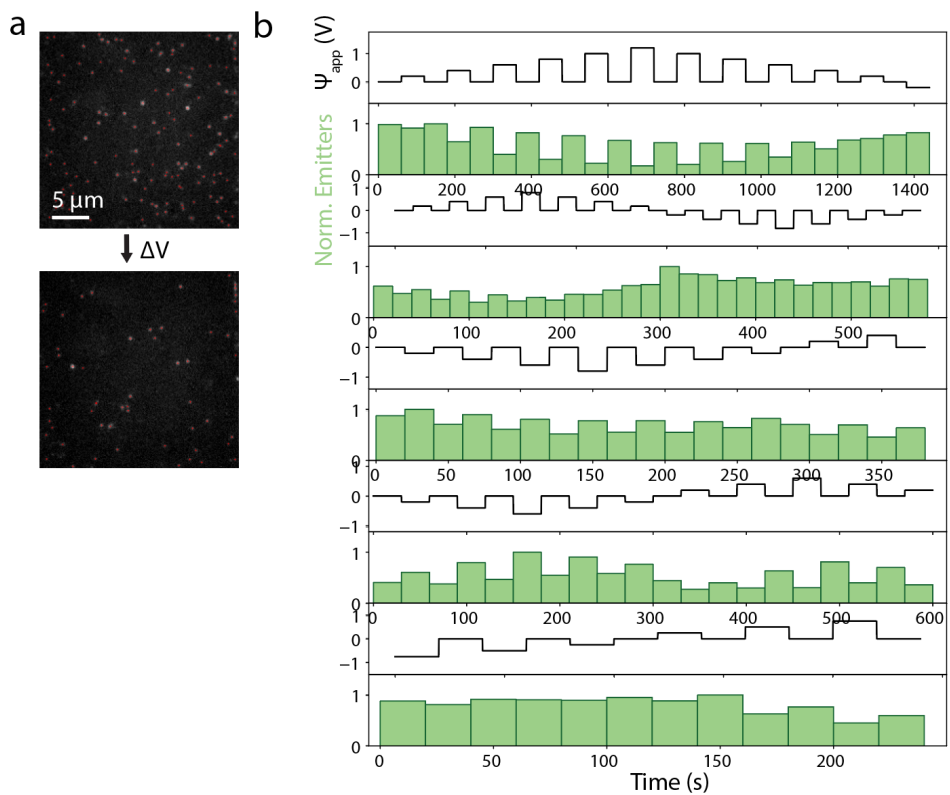

**Supplementary Figure 2: Quantification of electrochemical modulation in methanol**. **a)** Single molecule emitters are tracked and counted by localizing using ThunderSTORM, (see *Materials and Methods*). Two example frames taken with 50 ms acquisition time during a cycling experiment show these localizations marked in red. **b)** Square pulses of various patterns were applied to different flakes in methanol. The waveform is shown in black, and the corresponding binned localizations are shown in light green, below. Average changes from the same experiment's $\Psi_{app}$ = 0 V counts were used to characterize the response of methanol-hBN emitters to potential. All potentials are reported against Ag/AgCl.

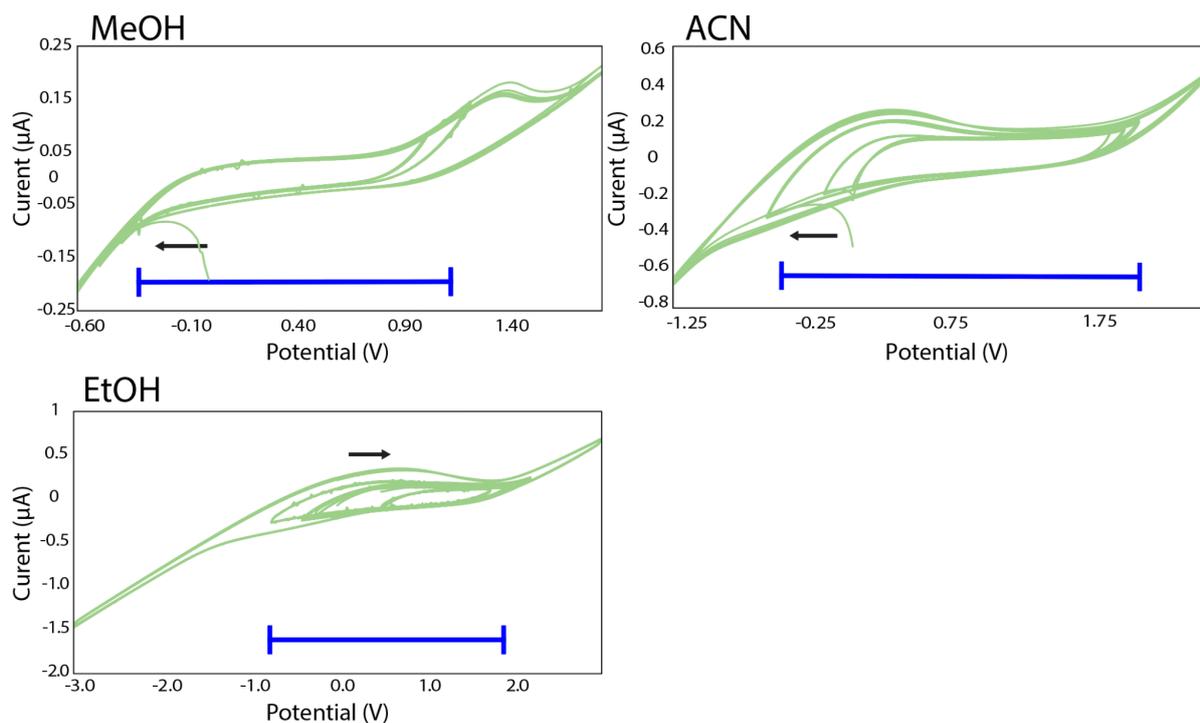

**Supplementary Figure 3: Characterizing inert regions of solvents and ITO via cyclic voltammetry.** Cyclic voltammetry was performed to determine the working potential range of solvents with an ITO working electrode vs Ag/AgCl leakless reference electrode in the three main solvents used in our experiments (methanol, acetonitrile, and ethanol). The stable working range is the flat region of the CV[1], shown by the blue bar and indicates the region wherein neither solvent nor electrode reacts. The scan rates were, 0.015 V/s, 0.05 V/s and 0.05 V/s for methanol, acetonitrile, and ethanol, respectively. The ITO coverslips have a diameter of 25 mm but only the center 8 mm diameter was exposed to solvent, corresponding to an active electrode surface area of 50 mm$^2$. The solvents were not used to quantify effects on hBN emitters with ITO outside these working regions.

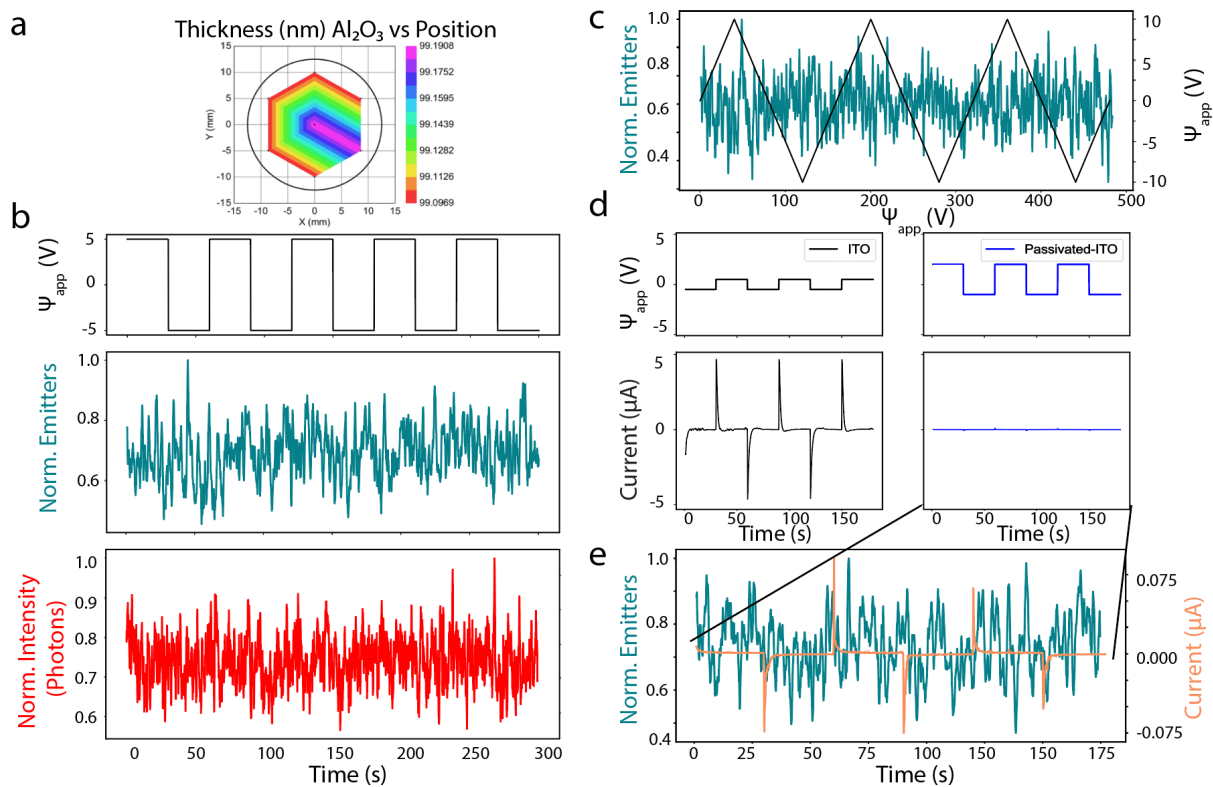

**Supplementary Figure 4: Alumina oxide passivation. a)** Ellipsometry confirms the 100 nm thick coating of alumina oxide deposited by atomic layer deposition over the ITO coverslip. **b)** In the absence of an active redox surface, the effect of cycling the voltage on emitter response was removed. A square voltage waveform applied to passivated ITO in methanol made no significant impact on the number of emitters on a flake (teal), even at high potentials (-/+5 V vs Ag/AgCl). There is also no modulation in average emitter intensity (red) per frame. **c)** A triangular waveform applied to a separate sample also shows no modulation of emitter counts even at very high potentials (-/+ 10 V, scan rate 250 mV/second). **d)** The un-passivated ITO currents (-/+0.5 V, black) compared to passivated ITO (-1 to +2 V, blue) in methanol are plotted on the same scale below the corresponding pulses. The magnitude of the passivated current is two orders of magnitude less than the un-passivated current, even for higher applied potentials. This indicates that, apart from minor currents, likely resulting from ion leakage through pores in $Al_2O_3$, the presence of the passivating layer effectively suppresses most reactions. This suppression seems to extend to our analyte reaction, as the effect of voltage cycling is effectively eliminated. **e)** The emitters are not modulated by the -1 to +2 V cycling (corresponding to the current trace above). The current from the bottom right panel of part d is scaled and can be seen to be less than 0.1 microamperes.

## Supplementary Discussion on Stray-Fields

To evidence that the source of optical modulation was an active redox surface rather than an orientation/electric field effect, we introduced a "stray-field" configuration wherein the ITO thin film electrode was replaced by a glass slide patterned with a thin titanium electrode and flakes were placed adjacent to the electrodes at different distances. Modulation was consistent with the out-of-plane measurements, despite the change in electrode substrate and orientation of the field. The strength of the signal was highly dependent on proximity of flake to electrode **(Supplementary Figure 5)**, indicating we had a spatially sensitive electrochemical sensor. Compared to the flake placed at 13 micrometers from the working electrode, the emitter signal modulation of the flake placed at 30 micrometers is considerably smaller. This characteristic indicates that the analyte species quenching emitters must diffuse from the electrode. However, the persistence of the effect at a distance of 30 microns also indicates that the bulk concentration of this analyte is small enough to be altered by the alternating currents.

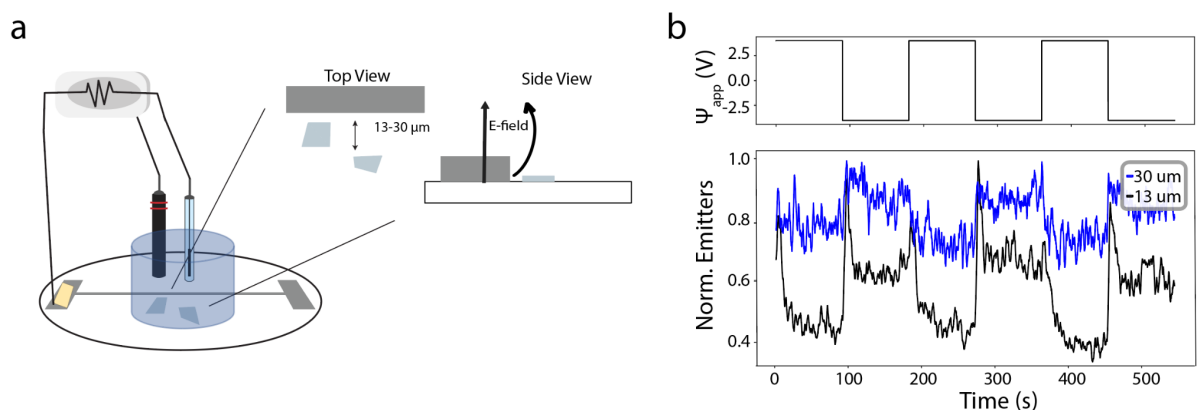

**Supplementary Figure 5: Stray-field measurements demonstrate distance-dependent modulation strength. a)** Schematic of the three-electrode configuration used to vary working electrode distance from flakes and apply stray electric field. Flakes were placed at a distance of 13 and 30 microns from the titanium working electrode. Glassy carbon and leakless Ag/AgCl were used as counter and reference electrodes, respectively which were connected to the potentiostat. **b)** The response of each flake to the same pulsed potential was monitored in acetonitrile. Although both flakes show the same behavior of decreased localizations at positive potentials, the magnitude is higher for the flake at 13 micrometers (black) than for the flake at 30 micrometers (blue) from the working electrode.

**Supplementary Discussion on Kinetics**

As discussed in the main text, the rate of reaction at the electrode is dependent on the electrochemical potential applied at the working electrode, k = k(V). The consistent and reversible modulation of analyte concentration is characterized in both reaction directions by exponential decay, indicative of a first order reaction. However, to evaluate the reaction order, emitter traces were fit to functions indicative of first (linear $Log[Analyte]\ vs\ Time$) and second ($[Analyte]^{-1}\ vs\ Time$) order reactions. In the context of simple linear regression, the coefficient of determination ($R^2$) evaluates how much of the variability of the y variable can be explained by variation in x. Here we have high fluctuations of emitter counts (even at $\Psi_{app}$=0 V), which explains the relatively low coefficients of determination. However, this does not impact our use of it to compare different models on the same data set. We also evaluated fits based on error and distributions of residuals over time (scatter plot) and residual distribution around 0 (histogram).

Based on our analysis, fitting to a first order reaction results in a higher goodness of fit and more evenly distributed residual values for the majority of potentials. **Supplementary Figure 6** shows an example pulse $\Psi_{app}$= 1.75 V but this was done for all potentials used in the kinetic analysis in **Figure 3**. It should be noted that calculating the kinetic constant for each electrochemical potential based on a second order reaction does result in the same trend as first order reaction fitting but errors are higher and the Tafel slope differs more from the current-derived value.

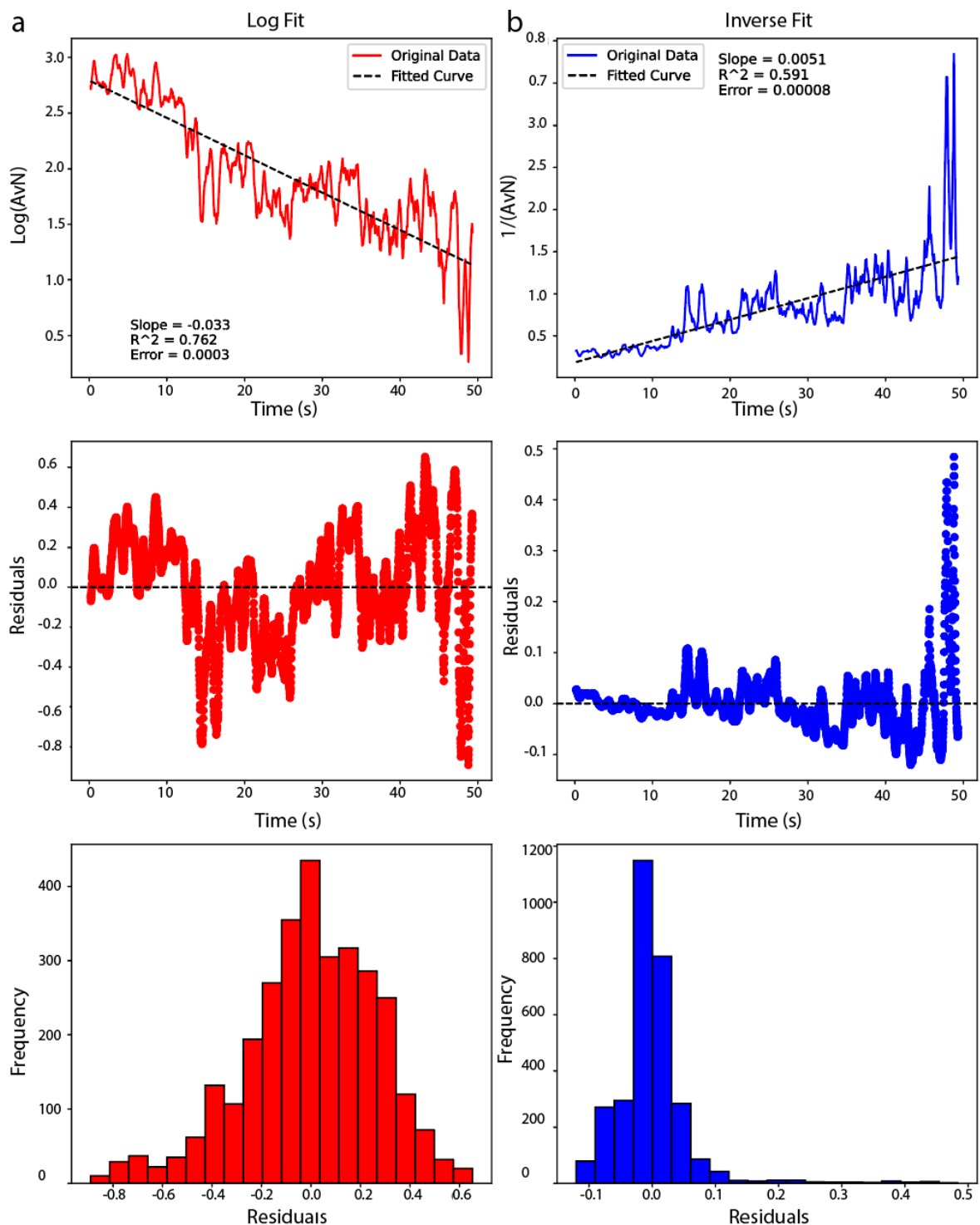

**Supplementary Figure 6: Comparison of first order and second order fitting for emitter count decays in response to positive potentials. a)** A first order reaction should have a linear $Log[Analyte]\ vs\ Time$ graph. Using the emitters as an optical readout for the quenching analyte, a linear fit was performed on $Log[Emitter]\ vs\ Time$ and was evaluated by $R^2$ and residual distributions. The example here shows emitter behavior for $\Psi_{app}$=1.75 V. **b)** Alternatively, a second order reaction should have a linear $[Analyte]^{-1}\ vs\ Time$ graph. Thus, a linear fit was also performed on $[Analyte]^{-1}\ vs\ Time$ and was evaluated by $R^2$ and residual distributions.

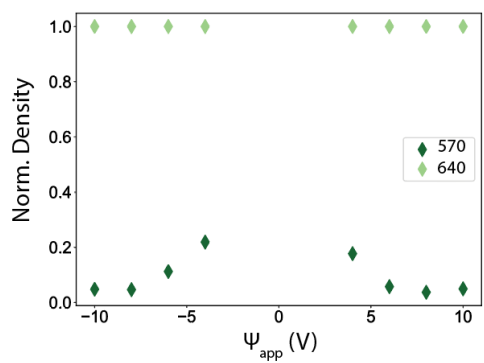

**Supplementary Figure 7: In-Plane spectral analysis**. At high electrochemical potentials the relative density of the secondary group of emitters (570 nm peak) is particularly reduced while the main peak around 640 nm is consistently the most prominent. This agrees with our findings in **Figure 3d-f**. Here we use titanium electrodes rather than ITO and the electrochemical potentials probed are higher in magnitude, but the surface is smaller.

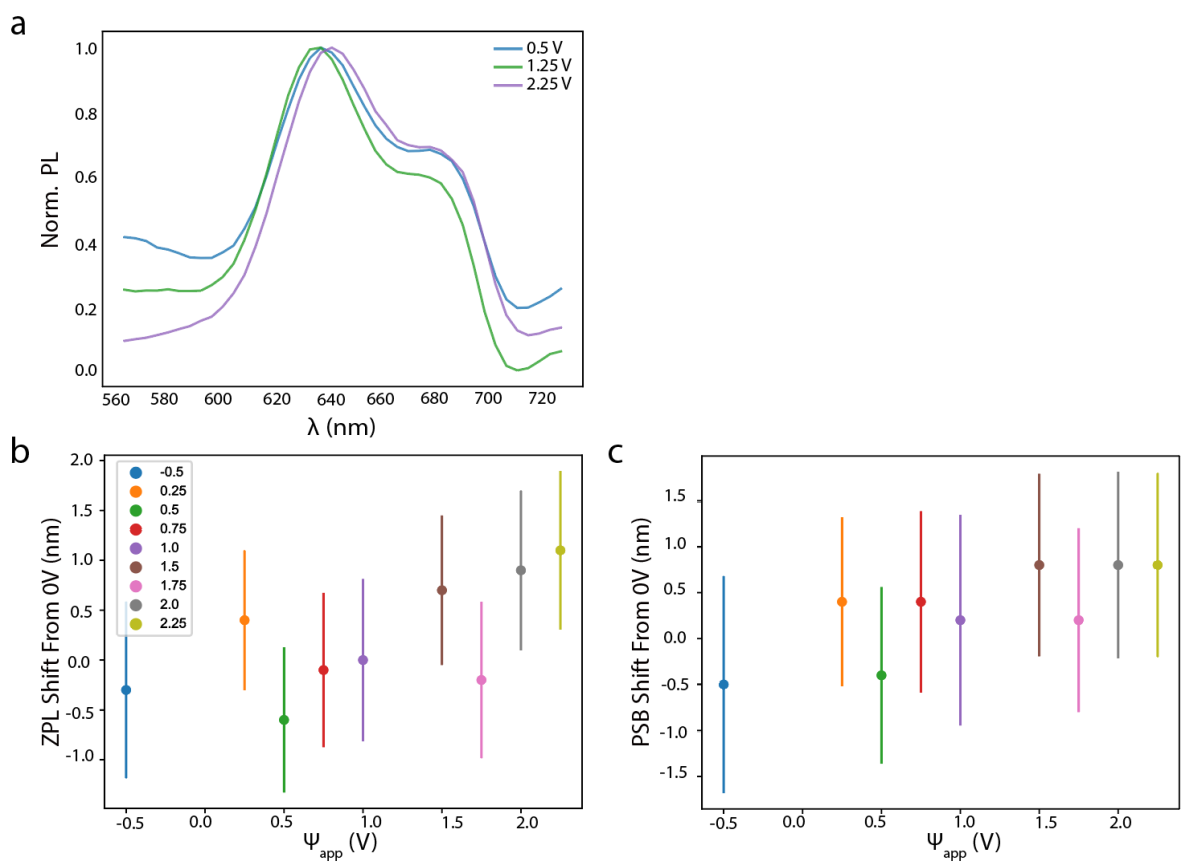

**Supplementary Figure 8: Out-of-plane averaged spectra shows no significant stark shift. a)** The ensemble-averaged spectrum of emitters measured at three different potentials in acetonitrile is presented. The averaging process differs from the use of individual emitter spectra, as depicted in **Figure 3d**, shown as a histogram and KDE. **b)** When fitting the averaged spectra to two Lorentzians (as described in *Materials and Methods*) the ZPL is obtained. Plotting ZPL shift from $\Psi_{app}$= 0 V for increasing potentials revealed no discernible trend. **c)** The PSB shift from 0 V also showed no significant trend as the electrochemical potential increased.

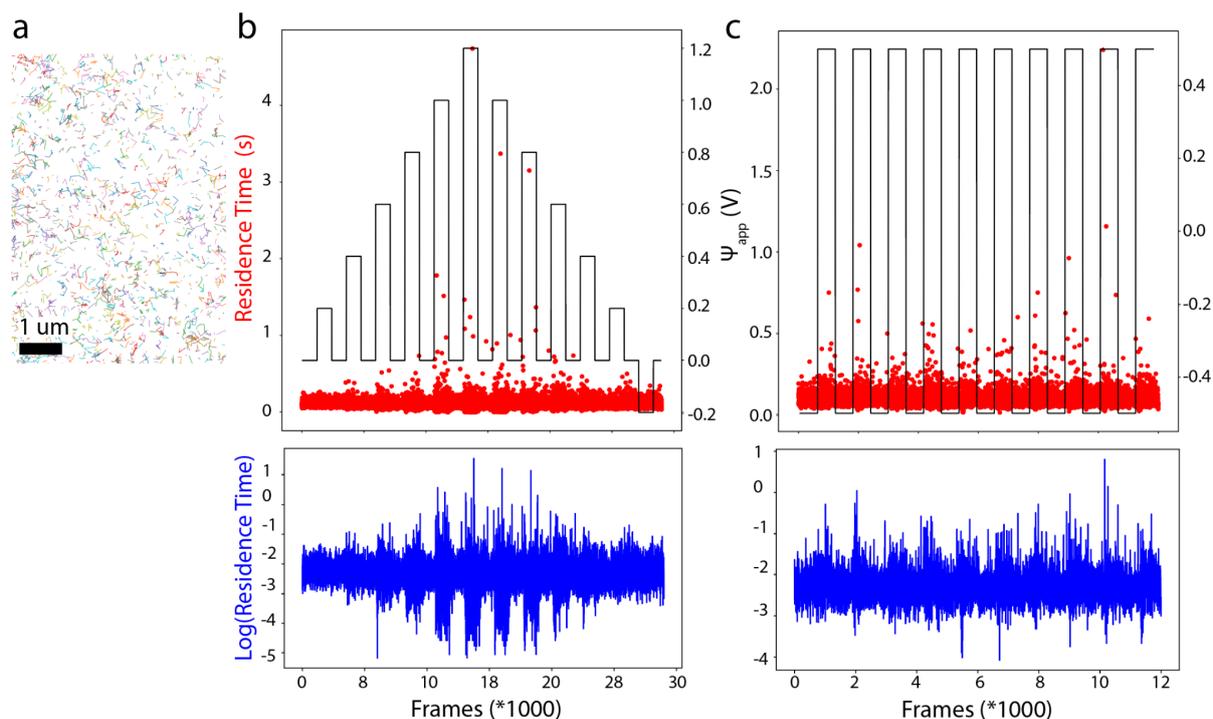

**Supplementary Figure 9: Residence time analysis. a)** By linking sub-diffraction localizations (ThunderSTORM) via single molecule tracking techniques (Trackpy Python implementation of Crocker-Grier algorithm) we determined the residence times of emitters. We defined the residence time as the number of subsequent frames an emitter is present from that frame forward. We characterize residence times by single trajectories, not single sites. Following a rationale detailed in previous work[2], we used a search range of 345 nm, adapted to our exposure time of 50 ms. Methanol-hBN surface emitter tracks at 0 V vs Ag/AgCl are shown after accumulating over 500 frames (50 ms per frame). **b)** A square waveform voltage ramp in methanol shows the changing distribution of residence times at increasingly positive potentials. The top panel just shows the average residence time for all emitters in that frame while the bottom panel shows the log of that average. The shape indicates that at positive voltages certain emitters are reduced from the beginning, which is intuitive with counts analysis in main text **Figures 1 and 2**. However, there is also an unexpected broadening of residence distribution, emitters that persist through the positive potentials seem to persist longer. **c)** The residence times (red) and log of residence times (blue) is also displayed for $\Psi_{app}$= +/-0.5 V in methanol and demonstrates the same trend.

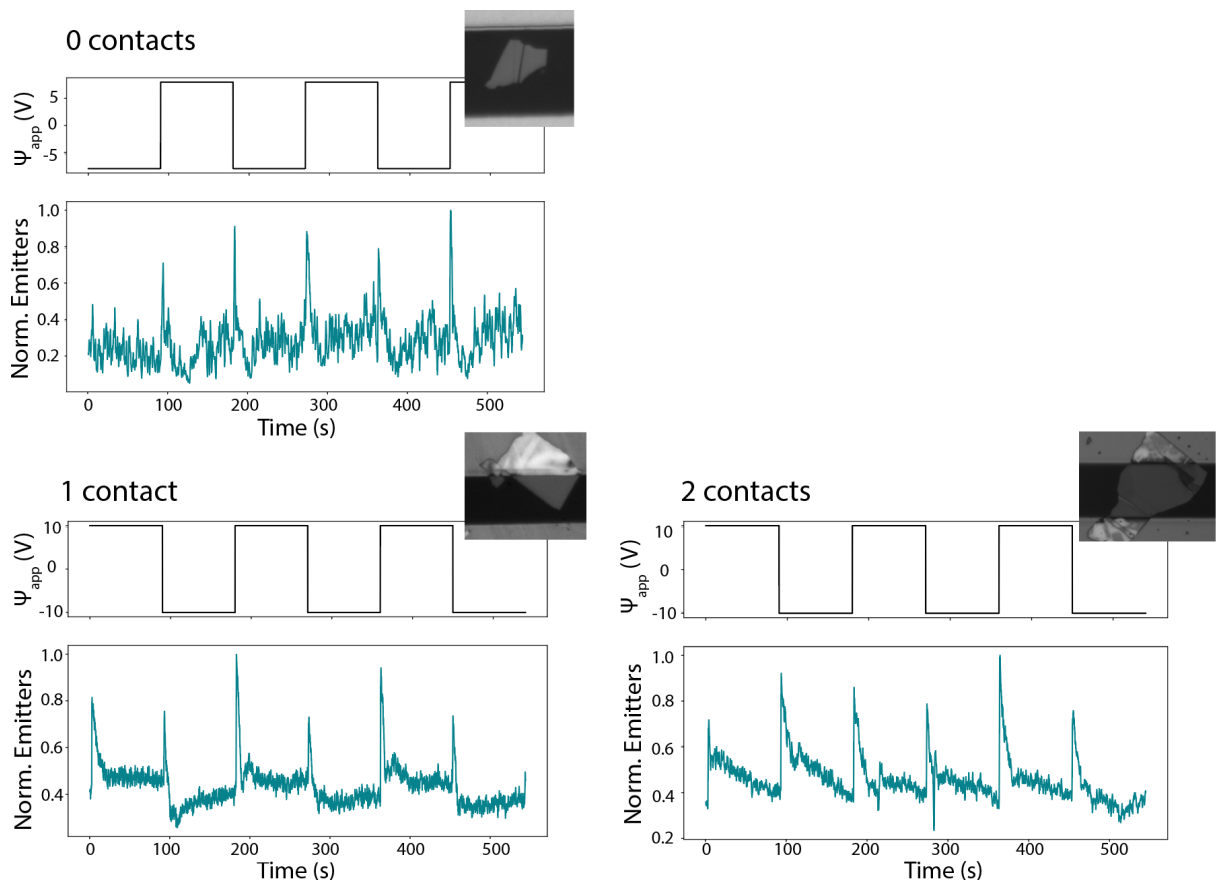

**Supplementary Figure 10: Contact configuration analysis.** We conducted experiments with various contact configurations to rule out the possibility of emitter modulation being caused by charge injection from the electrodes to hBN. Despite hBN being an insulator, we considered the possibility that high electrochemical potentials could impact defect charge states. Flakes were tested with no contact to either electrode, contact to one electrode, and contact to both electrodes. However, all configurations exhibited the same overall behavior. The variance in signal strength of these capacitive-looking switches can be understood as being due to larger distances between the electrode surface and the flake in the configurations with 0 contacts and 1 contact.

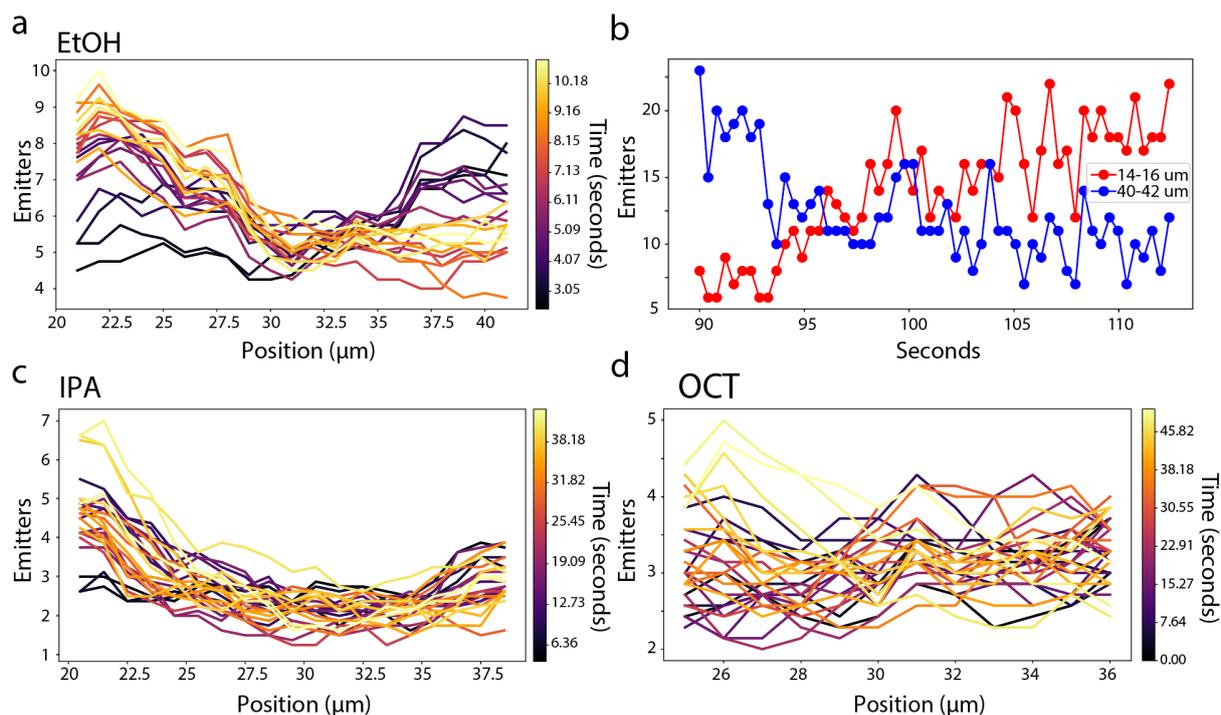

**Supplementary Figure 11: Monitoring concentration gradients in time. a)** The concentration gradient can be seen increasing over time after the switch in electrode polarization in the in-plane configuration with ethanol (EtOH). Switching the electrode polarization shows the switch in concentration gradient from high emitters on the right at the beginning to high on the left after 7 seconds. The time is reported from the time of the switch in electrode polarization. **b)** The change in local concentration for the same experiment in EtOH is also visualized by plotting the number of emitters in a 2 μm distance from the top electrode (red) and bottom electrode (blue). **c)** In isopropyl alcohol (IPA) we can also effectively monitor the concentration gradient buildup as the voltage is applied in the first cycle (not switching of polarization as in part a). The gradient takes about 3 times as long to build. **d)** Finally, in octanol (OCT), similar data is presented but the change in emitter density is considerably less, even after 45 seconds. Acetonitrile is shown in the main text (**Figure 4e**)

# Supplementary Discussion on Mechanisms

## Identifying and Excluding Candidate Reactant Species

Although we worked in the inert range of the electrodes and solvent (**Supplementary Figure 3**), products of an oxidation reaction of the organic solvent itself (including acetaldehyde, carboxylic acids, and $CO_2$) were considered as possible contributors to altered fluorescence. Intermediates and products of the bulk solvent were ultimately ruled out because the onset potential of the solvent reactions (indicated by curvature in the CVs) did not appear to coincide with the onset of our emitter modulation by electrochemical potential. Oxidation potentials depend heavily on the reaction conditions, such as pH, electrolyte, electrode type, etc., so it can be problematic to compare onset potentials between studies, especially since we were working in organic solvents without supporting electrolyte. Therefore, to understand the onset oxidation of a bulk solvent reaction in our system we generally refer to our CVs (**Supplementary Figure 3**). It has previously been shown that the onset of methanol oxidation with non-noble metals is generally >=1.2 V vs RHE[3] (>=1.397 V vs Ag/AgCl) and a current peak around this onset potential can be seen in our methanol CV. Thus, if the modulating species was an intermediate or product of this reaction, then the effect should begin around this point but the effect on emitters is seen at potentials much lower, even below 0.5 V vs Ag/AgCl in methanol. Furthermore, in ethanol the onset potential as seen from the CV is around 2 V vs Ag/AgCl, yet the effect is clearly seen on emitters at $\Psi_{app}$= +1 V (**Supplementary Figure 12**). Finally, since the current trace's curvature cannot be related to the effect on emitters, we are likely optically monitoring a reaction that is a subset to the overall current reported by the potentiostat.

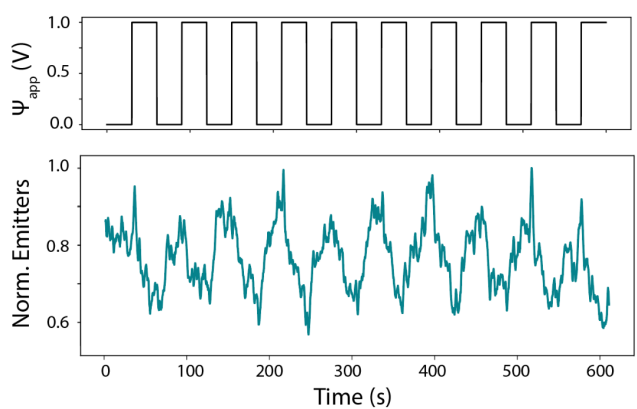

**Supplementary Figure 12: Ethanol cycling.** A square waveform was applied to the ITO electrode in ethanol in the three-electrode out-of-plane configuration to show the consistent behavior of emitters in other solvents. At +1 V the number of emitters decreases compared to 0V vs Ag/AgCl.

After eliminating bulk solvent reactions as the source of modulation, we sought to understand the activity at the electrode surface by examining the presence of trace species within the system and how their concentrations would respond to changes in electrochemical potential. For example, dioxygen would be present in the system from contact with the environment and its solubility in organic solvents[4]. This dioxygen could then be electrochemically reduced to oxygen radicals, present in trace amounts, which is a well-known quencher of fluorescent emitters[5]. However, this mechanism is inconsistent in that oxygen radicals are more likely to form at negative potentials while emitters in our system are instead quenched at positive potentials. Moreover, the extent of reaction is indicated by the extent of the change in center of mass, delta y, which relates directly to the solubility of water in the solvent (**Supplementary Figure 11**). Oxygen solubility increases with increasing chain length of the alcohols, but the gradient build-up is far lower in the long-chained octanol, indicating that oxygen is not the modulating species.

Other radicals were also considered, such as products of an oxidation reaction of the organic solvent: for example, alkoxy radicals may be formed at the electrode surface during oxidative cycling in alcohols but these radicals are highly reactive and would have short lifetimes (microseconds)[6], making it unlikely they would influence an electrode 30 μm away, such as is seen in the stray-field measurement (**Supplementary Figure 5**). Considering a diffusion constant on the order of magnitude of $10^{-9}$ m$^2$/s and assuming linear diffusion, diffusion of the radical from electrode to flake would take ~0.5 seconds, 5 orders of magnitude larger than the average lifetime of the alkoxy radicals.

## Experiments Showing the Involvement of Water and H⁺

Another trace species which would be present in all the solvents is water. Water and protons are known quenchers of red emitting dyes[5,7,8] and water would be present initially in the as-purchased solvents (100-200 ppm for acetonitrile and methanol and over 1000 ppm for ethanol)[9]. Water concentration would also be somewhat higher than as-purchased due to the exposure of the solvent to the environment–the cell was not air-tightly sealed during any experiment until anhydrous experiments (discussed in the following) and was not prepared in a glovebox.

To test the hypothesis that hBN emitters are quenched by interaction with protons, the concentration of which can be modulated by water oxidation, we deliberately introduced water (**Supplementary Figure 13**). As a result of introducing 3 M water into methanol, emitters were quenched by more than 50% and quenching percent was positively correlated with increasing concentration of water. We then directly introduced the proposed quencher, H⁺ (in the form of HCl), to methanol. This resulted in the same quenching trend as water but to a higher extent, reaching more than 50% quenching at only 70 mM, confirming that quenching was positively correlated with water and particularly H⁺.

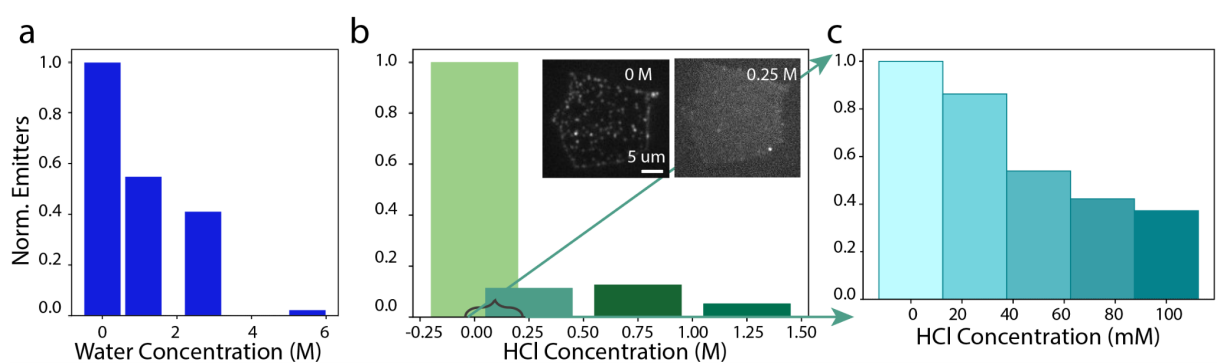

**Supplementary Figure 13: Deliberate introduction of quenchers. a)** The average number of emitters per frame for a single flake for 2400 frames is plotted against increasing water concentration. Emitters per frame is normalized. As water concentration increases the concentration of emitters in methanol decreases. At 2.5 M the emitter counts are decreased by more than 60%. **b)** The average number of emitters per frame is plotted in for increasing concentration of hydrochloric acid. The concentration was increased by mixing pure methanol with a stock of 5 M HCl in methanol, in which the HCl had been bubbled in. The inset images show the flake imaged in pure methanol vs with 250 mM HCl. **c)** A lower concentration range was tested using a fresh sample. This showed that emission is quenched by more than 50% at only 70 mM.

We then considered that removing the reactant candidate (water) should eliminate the modulation of the quencher concentration. To remove adventitious water, we used an activated 3Å molecular sieve over 48 hours in a glovebox to dry acetonitrile. The water content in acetonitrile after this treatment should be imperceptible[9]. The acetonitrile was then loaded in ambient conditions into an electrochemical chamber. The electrochemical chamber was sealed from the environment and cycling experiments were performed and their results, referred to as "dehydrated" were compared to the results from normal acetonitrile, referred to as "wet". Consistent with our proposal, this drying removed the effect of voltage cycling when switching between (-/+0.5 V) whereas wet acetonitrile has a strong response to this pulse (**Supplementary Figure 14**). When the magnitude of the pulses increased, an effect on emitter counts was seen even in the dry acetonitrile but still to a lesser extent, indicating that water plays an important role in our modulating reaction mechanism. One should note that, due to technical limitations, the experiment (including the chamber assembly) was not performed in a glovebox, so the introduction of water by contact with the environment was possible during this time.

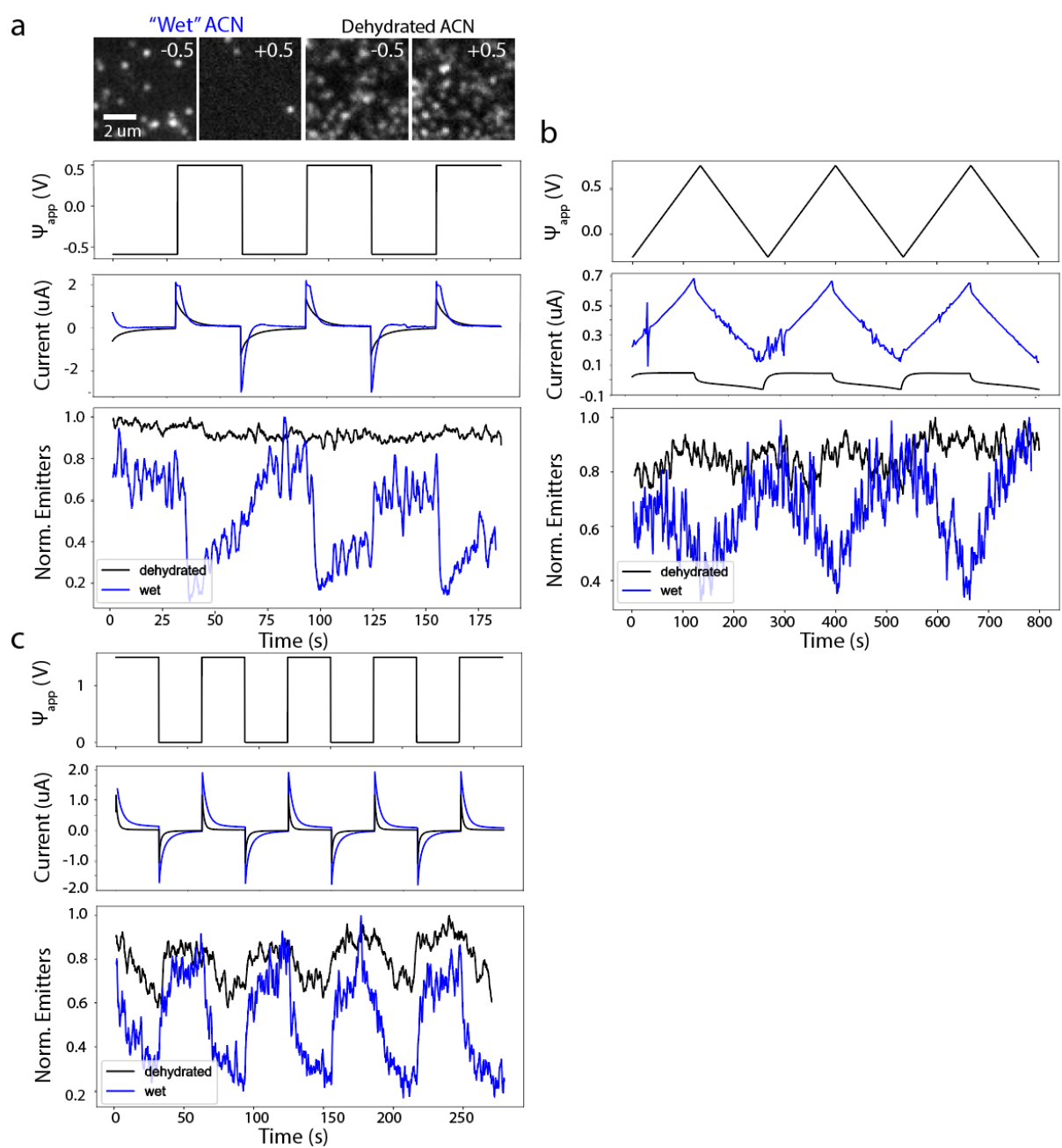

**Supplementary Figure 14: Deliberate Removal of Water. a)** Example frames show the visible difference in modulation extent between wet and dehydrated acetonitrile when cycling between $\Psi_{app}$= -/+0.5 V at the ITO electrode in the three-electrode out-of-plane configuration. The corresponding applied waveform, currents, and normalized emitters per frame are shown below. Wet acetonitrile is shown in blue while dehydrated acetonitrile is shown in black. Dehydration effectively eliminates modulation by cycling and decreases the current. **b)** A triangular waveform was applied to the ITO working electrode in the dehydrated acetonitrile and the currents and normalized emitters per frame are shown in comparison to the normal, wet, acetonitrile. Modulation of emitters by electrochemical potential is again effectively eliminated. **c)** Finally, the effect of higher applied electrochemical potential ($\Psi_{app}$ = + 1.25) recovers the impact on dry acetonitrile although the effect is still lower than in wet acetonitrile. It should also be noted that the experiments were done in the order of presentation and any leakage of humidity through our sealing would occur over time.

From the modulations of water and $H^+$ concentrations in our solution we can confirm that these species play an important role in emitter quenching. We thus propose that the emitter density is modulated by the electrochemical reduction/oxidation of water which results in changing concentration of $H^+$. In our in-plane experiments, oxidation and reduction of $H^+$ can occur in the same field of view, explaining the increase of emitters on one side and decrease on the other. This is schematically illustrated in **Supplementary Figure 15**.

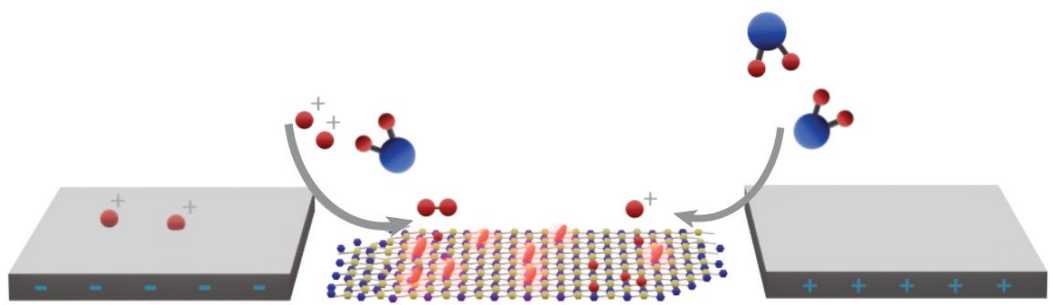

**Supplementary Figure 15: Interpretation of modulation mechanism.** An interpretation of the system that was shown optically in **Figure 4**. In this interpretation the modulation is due to the reduction/oxidation of $H^+$ and water, respectively. $H^+$ can be catalytically reduced to $H_2$ at the surface of the titanium electrode or recombined into water using molecular oxygen. Alternatively, at the positive electrode, water is oxidized to form $H^+$, which can quench emitters. The emitters then appear at higher density near the negative electrode whereas diminished fluorescence is observed near the positive electrode.

## Proposed Electrochemical Mechanism

We propose that the modulation of emitters during electrochemical cycling results from the reduction/oxidation reactions of water, which changes the concentration of $H^+$, ultimately quenching the emitters. The oxidation reaction at positive potentials is described by electrolysis, the electrochemical splitting of liquid water into hydrogen and oxygen using electricity[10]. Meanwhile, at the cathode, higher emitter density is recorded as $H^+$ is consumed (e.g. reduced to $H_2$ or recombined into water). One should note that the transient behavior of emitters is not induced by adding water or introducing electrochemical cycling but is present in all experiments on hBN organic solvent emitters, as previously reported[2].

As was discussed briefly in the main text, the increased reactivity of water in acetonitrile can be understood by the alteration of its molecular aggregation behavior, with its oxidation potential shown to steadily decrease with decreasing water concentrations[11]. However, just as our redox modulation is not unique to acetonitrile, the isolation of water molecules in mixtures has also been observed and modeled with ionic liquids/water[12] and methanol/water[13], respectively. Moreover, in our experiments we have the addition of irradiating light, which has been seen to increase reactivity and drive the water splitting reaction[14,15].

Finally, the development of $H^+$ sensors based on fluorescence is valuable to the innovation of *in-operando* probes. In our system, hBN organic solvent emitters were shown to be highly sensitive to $H^+$ concentration in methanol, the monitoring of which is decidedly relevant to development of methanol fuel cells[16,17]. The spatiotemporal resolution of our system also means that the $H^+$ concentration could be monitored as a function of distance from electrode or membrane over time.

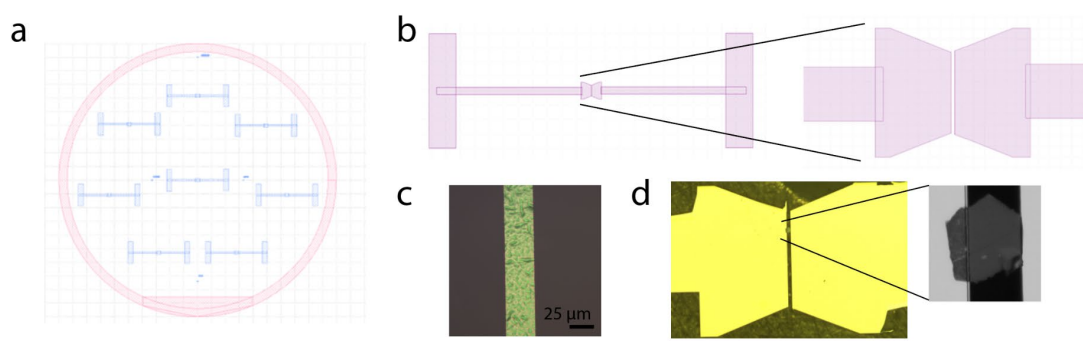

**Supplementary Figure 16: Wafer for in-plane electrode fabrication. a)** In-plane and stray-field electrodes are patterned onto 8 coverslips at a time using E-beam evaporation with a $Si_3N_4$ shadow mask. **b)** The stencil mask was created using direct laser writing and positive photolithography. Thus, everywhere shown in pink was etched away and a thin nitride membrane remains. The distance between electrodes was varied by controlling the width of the remaining nitride membrane. **c)** The optical microscope image shows an example nitride membrane that is 30 μm wide. **d)** The result of evaporating 100-250 nm of titanium onto a 25 mm diameter glass coverslip using the stencil mask is seen. A flake has been transferred deterministically between the electrodes using a PDMS stamp.

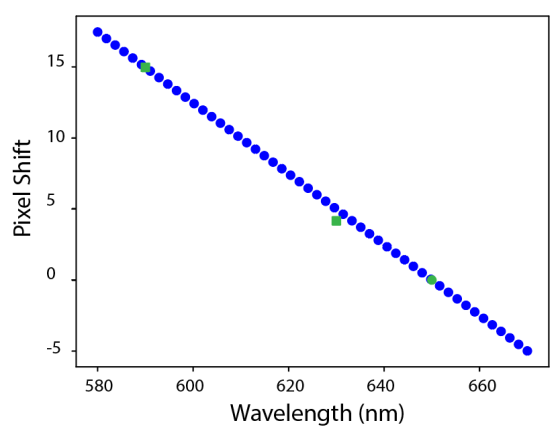

**Supplementary Figure 17: Bead calibration of spectral SMLM.** Linear calibration for the vertical shift in the spectral channel as a function of emission wavelength was done using broadband beads and narrow bandpass filters. On the spectral path there is a dispersive prism that leads to an approximately

linear shift in the position of the emitter relative to the wavelength. Recall from the *Materials and Methods* section that $(x_{SPEC}, y_{SPEC}) = A \times (x_{LOC}, y_{LOC}) + B$ where A is a 2 × 2 matrix and B is a column vector. The B vector was determined via this linear calibration. This sketch is of the same method previously reported[18] with updated calibration.

# Captions For Supplementary Videos

Videos are available on Zenodo at: https://doi.org/10.5281/zenodo.11206276

**Supplementary Video 1: Out-of-plane emitter control.** The video included is a wide-field view of an hBN flake immersed in acetonitrile while an electrochemical potential of the ITO working electrode is cycled in the three-electrode out-of-plane configuration between +1.5 V and 0 V vs Ag/AgCl. The change in potential induces a change in density of emitters. This data corresponds to a flake used in spectral analysis (data presented in **Fig. 3c–e**). Continuous ~1.6 kW cm$^{-2}$ illumination with a 561 nm laser is used. The original images were acquired at a rate of 50.091 ms per frame but here we combined frames to present a lighter video with a lower sampling rate (1 s). The total experiment took 750 seconds, but we present the image stack at a rate of 10 frames per second to make the video only 75 seconds. The scale bar is 5 μm.

**Supplementary Video 2: In-plane emitter control.** Here we present a wide-field video of the hBN flake shown in **Figure 4b-d** in between two titanium electrodes. Continuous ~1 kW cm$^{-2}$ illumination is used with a 561 nm laser while the polarization of the electrodes is cycled in the two-electrode in-plane configuration, inducing a change in density of emitters correlated with potential. The high-density region follows the negatively charged electrode. The original images were acquired with a 50.091 ms rate, but here we combined frames to present a lighter video with a lower sampling rate (2.5 s). The total experiment took 540 seconds, but we display 5 of the stacked images per second, making make the video around 40 seconds. The scale bar is 5 μm.

# Supplementary References